\documentclass[aps, prl,reprint,superscriptaddress,amssymb]{revtex4-1}

\usepackage{amsmath}

\usepackage{graphicx}
\usepackage{floatrow}
\usepackage[leftcaption]{sidecap}
\sidecaptionvpos{figure}{right}
\usepackage{amsmath,amsfonts,amsbsy,amssymb,array,accents,dsfont}
\usepackage{enumerate,array,latexsym,graphicx,mathrsfs,verbatim,psfrag}
\usepackage{bm} 
\usepackage[normalem]{ulem}
\usepackage{multirow}
\usepackage{cancel}
\usepackage{chngpage} 

\usepackage{booktabs} 
\usepackage[usenames]{color}
\usepackage[utf8]{inputenc}
\usepackage[english]{babel}
\usepackage{fancybox}
\usepackage[dvipsnames]{xcolor}

\usepackage{tikz}
\usepackage{pgfplots}
\usetikzlibrary{decorations.text,intersections, pgfplots.fillbetween}
\usetikzlibrary{math}
 \usetikzlibrary{snakes,3d,shapes.geometric,shadows.blur}
\usepackage{tikz-3dplot}

\usetikzlibrary{positioning}
\usetikzlibrary{automata}
\usetikzlibrary{arrows}
\usetikzlibrary{calc}
\usetikzlibrary{decorations.markings}
\usetikzlibrary{decorations.pathreplacing}
\usetikzlibrary{intersections}
\usetikzlibrary{positioning}
\usetikzlibrary{topaths}
\usetikzlibrary{shapes.geometric}
\usetikzlibrary{shapes.misc}
\tikzset{cf-group/.style = {
    shape = rounded rectangle, minimum size=1.0cm,
    rotate=90,
    rounded rectangle right arc = none,
    draw}}
\tikzset{cross/.style={path picture={ 
  \draw[black]
(path picture bounding box.south east) -- (path picture bounding box.north west) (path picture bounding box.south west) -- (path picture bounding box.north east);
}}}

\tikzset{unode/.style=
{black, circle,draw,thick,fill=black!100 ,minimum size=1mm}}

\tikzset{sunode/.style=
{black, circle,draw,thick,fill=yellow!100 ,minimum size=1mm}}

\tikzset{fnode/.style=
{black, rectangle,draw,thick,minimum size=1mm}}

\tikzset{afnode/.style=
{blue,rectangle,draw,thick,minimum size=1mm}}

\usepackage{sidecap}

\sidecaptionvpos{figure}{c}

\usepackage{mciteplus}

\usepackage[colorlinks=true,      linkcolor=blue,      urlcolor=blue,      
            filecolor=blue,      citecolor=blue,       pdfstartview=FitH,     
						pdfpagemode=UseNone,      bookmarksopen=true]{hyperref}  
\usepackage[all]{hypcap}     

\newcommand{\be}{\begin{equation}}
\newcommand{\ee}{\end{equation}}
\newcommand{\ba}{\begin{array}}
\newcommand{\ea}{\end{array}} 
\newcommand{\bi}{\begin{itemize}}
\newcommand{\ei}{\end{itemize}}
\def\vec#1{\bm{#1}}
\def\bea#1\eea{\allowdisplaybreaMs \begin{align}#1\end{align}}
 \newcommand{\ben}{\begin{enumerate}}
\newcommand{\een}{\end{enumerate}}
\newcommand{\bean}{\begin{eqnarray*}}
\newcommand{\eean}{\end{eqnarray*}}
\newcommand{\eref}[1]{(\ref{#1})}

\newcommand{\nn}{\nonumber}

\newcommand{\tr}{\mathrm{Tr}}

\newcommand{\CT}{{\cal T}}
\newcommand{\CD}{{\cal D}}

\newcommand{\CO}{{\cal O}}

\newcommand{\CN}{{\cal N}}

\newcommand{\frsu}{\mathfrak{su}}
\newcommand{\fru}{\mathfrak{u}}

\newcommand{\frg}{\mathfrak{g}}

\newcommand{\wt}{\widetilde}

\newcommand{\ch}{\cosh \pi}
\newcommand{\s}{\sigma}

\newcommand{\Tabref}[1]{Table~\ref{#1}}

\newcommand{\Figref}[1]{Figure~\ref{#1}}
\newcommand{\figref}[1]{Fig.~\ref{#1}}
\renewcommand{\eqref}[1]{(\ref{#1})}

\usepackage{titlesec}
\titlespacing*{\section}
{0pt}{0.5cm}{0.2cm}
\titlespacing*{\subsection}
{0pt}{0.25cm}{0.1cm}

\begin{document}

\title{Emergent Global Symmetry from IR N-ality}

\author{Anindya Dey}
\email{anindya.hepth@gmail.com}
\affiliation{Department of Physics and Astronomy, Johns Hopkins University, 3400 North Charles Street, Baltimore, MD 21218, USA}

\begin{abstract}

\noindent We present a new family of IR dualities in three space-time dimensions with eight supercharges. 
In contrast to 3d mirror symmetry, these dualities map Coulomb branches to Coulomb branches and Higgs branches 
to Higgs branches in the deep IR. 
For a large class of quiver gauge theories with an emergent Coulomb branch global symmetry, 
one can construct a sequence of such dualities by step-wise implementing a set of quiver mutations. 
The duality sequence gives a set of quiver gauge theories which flow to the same IR 
fixed point --  a phenomenon we refer to as IR N-ality. We show that this set of N-al quivers 
always contains a theory for which the rank of the IR Coulomb branch symmetry is manifest in the UV. 
For a special subclass of theories, the emergent symmetry algebra itself can be read off from the quiver description of the 
aforementioned theory. 

\end{abstract}

\maketitle


\noindent \textit{Introduction.} Some of the most interesting non-perturbative phenomena in QFTs
in three and four space-time dimensions \cite{Seiberg:1994pq, Aharony:1997gp, Giveon:2008zn, Intriligator:1996ex} 
arise in the IR limit, where the theories may become  
strongly-interacting at special points of the vacuum moduli space. Broadly speaking, the properties of a QFT  that arise in the neighborhood 
of such special points but are not manifest in the UV description, are collectively referred to as \textit{emergent}
properties. A particularly important example involves the global symmetry of the QFT at these special points. 

3d $\CN=4$ theories provide a rich laboratory for studying non-perturbative phenomena in 
QFTs. The theories are super-renormalizable in the UV and generically flow to strongly-coupled SCFTs in the IR. 
The vacuum moduli space has two distinguished branches : the Higgs branch (HB), which is protected from quantum corrections 
by a non-renormalization theorem, and the Coulomb branch (CB), which receives 1-loop as well as non-perturbative corrections. 
We will focus on theories which are \textit{good} in the Gaiotto-Witten sense \cite{Gaiotto:2008ak} -- the two branches in this case 
intersect at a single point where the IR SCFT lives. 3d $\CN=4$ theories also present interesting examples of IR duality -- a pair of distinct theories in 
the UV flowing to the same IR SCFT. A particularly important example of such a duality is 3D Mirror Symmetry \cite{Intriligator:1996ex, deBoer:1996mp} 
which acts by mapping the CB of one theory to the HB of the other and vice-versa, in the deep IR.

The HB 0-form symmetry, including its global form, is classically manifest. For the CB, however, the IR symmetry algebra $\frg^{\rm IR}_{\rm C}$ may be larger 
compared to the UV-manifest symmetry $\frg^{\rm UV}_{\rm C}$. If the rank of the IR symmetry is greater than the UV-manifest rank, we will refer to the IR symmetry as 
\textit{emergent}, otherwise we will simply refer to it as \textit{enhanced}. 

\begin{figure}[htbp]
\begin{center}
\scalebox{0.55}{\begin{tikzpicture}
\node[] (100) at (-3,0) {};
\node[] (1) at (-1,0) {};
\node[unode] (2) at (0,0) {};
\node[text width=.2cm](31) at (0.1,-0.5){$N_1$};
\node[unode] (3) at (2,0) {};
\node[text width=.2cm](32) at (2.1,-0.5){$N_2$};
\node[] (4) at (3,0) {};
\node[] (5) at (4,0) {};
\node[unode] (6) at (5,0) {};
\node[text width=.2cm](33) at (5.1,-0.5){$N_{\alpha}$};
\node[unode] (7) at (7,0) {};
\node[text width=.2cm](34) at (7.1,-0.5){$N_{\alpha+1}$};
\node[unode] (8) at (9,0) {};
\node[text width=.2cm] (40) at (9.1,-0.5) {$N_{\alpha+2}$};
\node[fnode] (9) at (9,-2) {};
\node[text width=.2cm] (40) at (9.5,-2) {$M_{\alpha+2}$};
\node[] (10) at (10,0) {};
\node[] (11) at (12,0) {};
\node[fnode] (20) at (0,-2) {};
\node[fnode] (21) at (2,-2) {};
\node[fnode] (22) at (5,-2) {};
\node[fnode] (27) at (7,-2) {};
\node[text width=.2cm](23) at (0.5,-2){$M_1$};
\node[text width=.2cm](24) at (2.5,-2){$M_2$};
\node[text width=.2cm](25) at (5.5,-2){$M_\alpha$};
\node[text width=.2cm](26) at (7.5,-2){$M_{\alpha +1}$};
\draw[thick] (1) -- (2);
\draw[thick] (2) -- (3);
\draw[thick] (3) -- (4);
\draw[thick,dashed] (4) -- (5);
\draw[thick] (5) -- (6);
\draw[thick] (6) -- (7);
\draw[thick] (7) -- (8);
\draw[thick] (7) -- (27);
\draw[thick] (8) -- (9);
\draw[thick] (8) -- (10);
\draw[thick,dashed] (10) -- (11);
\draw[thick,dashed] (1) -- (100);
\draw[thick] (2) -- (20);
\draw[thick] (3) -- (21);
\draw[thick] (6) -- (22);
\end{tikzpicture}}
\end{center}
\caption{\footnotesize{A linear quiver with unitary gauge nodes. A black circular node with label $N$ represents a $U(N)$ gauge node, a black square node 
with label $F$ represents $F$ hypermultiplets in the fundamental representation, and a thin black line connecting two gauge nodes is a bifundamental hypermultiplet.}}
\label{fig: LQgen}
\end{figure}
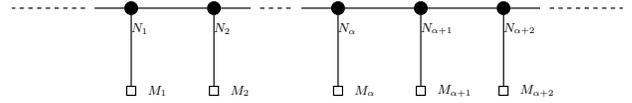

A very well-known example of a CB symmetry enhancement involves a linear quiver gauge theory with unitary gauge nodes, as shown in \figref{fig: LQgen}. The theory is good in the Gaiotto-Witten sense \cite{Gaiotto:2008ak} 
if the integers $e_\alpha = N_{\alpha-1}+N_{\alpha+1} + M_\alpha - 2N_\alpha$ (balance parameter for the $\alpha$-th node) obey the condition 
$e_\alpha \geq 0, \forall \alpha$.

For every unitary gauge node, there exists a $\fru(1)$ topological symmetry, and the CB global symmetry manifest in the UV is simply 
$\frg^{\rm UV}_{\rm C}=\oplus^L_{\alpha=1}\, \fru(1)_\alpha$. The UV-manifest rank is $\text{rk}(\frg^{\rm UV}_{\rm C})=L$, 
where $L$ is the total number of gauge nodes. In the IR, every array of $k$ consecutive balanced (i.e. $e_\alpha=0$) gauge nodes contributes an $\frsu(k+1)$ factor to the 
symmetry algebra, while every overbalanced node (i.e. $e_\alpha > 0$) contributes a factor of $\fru(1)$\cite{Gaiotto:2008ak}. The IR global symmetry algebra therefore has the generic form:
\be \label{LQ-IRsymm}
\frg^{\rm IR}_{\rm C} = \oplus_{\alpha}\, \frsu(k_\alpha +1)_\alpha  + \oplus_{\beta}\, \fru(1)_\beta,
\ee
where $\alpha$ labels every array of $k_\alpha$ consecutive balanced gauge nodes, while $\beta$ labels the overbalanced nodes. Note that, while 
$\frg^{\rm IR}_{\rm C} \neq \frg^{\rm UV}_{\rm C}$, we have $\text{rk}(\frg^{\rm IR}_{\rm C})=\text{rk}(\frg^{\rm UV}_{\rm C})=L$. Therefore, the rank of the IR global symmetry is manifest in the UV. For every $\fru(1)$ factor in $\frg^{\rm UV}_{\rm C}$, one can turn on a triplet of Fayet-Iliopoulos (FI) parameters in the UV Lagrangian. In the IR, 
these parameters account for $\CN=4$-preserving mass deformations of the SCFT, deforming/resolving the HB.  

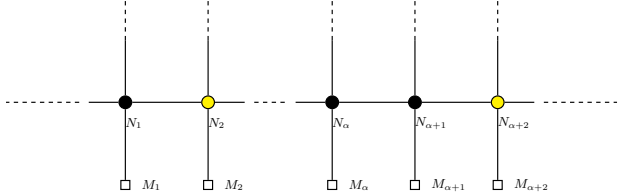
\begin{figure}[htbp]
\begin{center}
\scalebox{0.55}{\begin{tikzpicture}
\node[] (100) at (-3,0) {};
\node[] (1) at (-1,0) {};
\node[unode] (2) at (0,0) {};
\node[text width=.2cm](31) at (0.1,-0.5){$N_1$};
\node[sunode] (3) at (2,0) {};
\node[text width=.2cm](32) at (2.1,-0.5){$N_2$};
\node[] (4) at (3,0) {};
\node[] (5) at (4,0) {};
\node[unode] (6) at (5,0) {};
\node[text width=.2cm](33) at (5.1,-0.5){$N_{\alpha}$};
\node[unode] (7) at (7,0) {};
\node[text width=.2cm](34) at (7.1,-0.5){$N_{\alpha+1}$};
\node[sunode] (8) at (9,0) {};
\node[text width=.2cm] (40) at (9.1,-0.5) {$N_{\alpha+2}$};
\node[fnode] (9) at (9,-2) {};
\node[text width=.2cm] (40) at (9.5,-2) {$M_{\alpha+2}$};
\node[] (10) at (10,0) {};
\node[] (11) at (12,0) {};
\node[fnode] (20) at (0,-2) {};
\node[fnode] (21) at (2,-2) {};
\node[fnode] (22) at (5,-2) {};
\node[fnode] (27) at (7,-2) {};
\node[text width=.2cm](23) at (0.5,-2){$M_1$};
\node[text width=.2cm](24) at (2.5,-2){$M_2$};
\node[text width=.2cm](25) at (5.5,-2){$M_\alpha$};
\node[text width=.2cm](26) at (7.5,-2){$M_{\alpha +1}$};
\draw[thick] (1) -- (2);
\draw[thick] (2) -- (3);
\draw[thick] (3) -- (4);
\draw[thick,dashed] (4) -- (5);
\draw[thick] (5) -- (6);
\draw[thick] (6) -- (7);
\draw[thick] (7) -- (8);
\draw[thick] (7) -- (27);
\draw[thick] (8) -- (9);
\draw[thick] (8) -- (10);
\draw[thick,dashed] (10) -- (11);
\draw[thick,dashed] (1) -- (100);
\draw[thick] (2) -- (0,1.5);
\draw[thick, dashed] (0,1.5) -- (0,2.5);
\draw[thick] (2,1.5) -- (3);
\draw[thick, dashed] (2,1.5) -- (2,2.5);
\draw[thick] (5,1.5) -- (6);
\draw[thick, dashed] (5,1.5) -- (5,2.5);
\draw[thick] (7,1.5) -- (7);
\draw[thick, dashed] (7,1.5) -- (7, 2.5);
\draw[thick] (9,1.5) -- (8);
\draw[thick,dashed] (9,1.5) -- (9,2.5);
\draw[thick] (2) -- (20);
\draw[thick] (3) -- (21);
\draw[thick] (6) -- (22);
\end{tikzpicture}}
\end{center}
\caption{\footnotesize{A generic quiver with unitary/special unitary gauge nodes with at least one of the $SU$ nodes being balanced. A yellow circular node with label $N$ represents an $SU(N)$ gauge node.}}
\label{fig: USUgen}
\end{figure}

More generally, however, one may have $\text{rk}(\frg^{\rm IR}_{\rm C}) > \text{rk}(\frg^{\rm UV}_{\rm C})$, which implies that some of the mass deformations 
are simply not visible in the UV Lagrangian. These are often referred to as theories with ``hidden FI parameters" \cite{Hanany:1997gh, Feng:2000eq, Gaiotto:2008ak}.
A particularly interesting class is given by quiver gauge theories with unitary and special unitary gauge nodes and hypermultiplets in the fundamental/bifundamental representations (see \figref{fig: USUgen}), with at least one of the special unitary nodes being \textit{balanced} i.e. the total number of fundamental/bi-fundamental hypers associated with a given $SU(N_\alpha)$ node is $2N_\alpha-1$. The latter condition ensures that the quiver has an emergent IR CB symmetry, as we will see momentarily. We will restrict our discussion to quivers defined by tree-graphs. Quivers with loops, which can introduce additional emergent symmetry, will be discussed elsewhere.

\begin{figure}[htbp]
\begin{center}
\scalebox{0.5}{\begin{tikzpicture}
\node[] (100) at (-3,0) {};
\node[] (1) at (-1,0) {};
\node[unode] (2) at (0,0) {};
\node[text width=.2cm](31) at (0.1,-0.5){$N_1$};
\node[unode] (3) at (2,0) {};
\node[text width=.2cm](32) at (2.1,-0.5){$N_2$};
\node[] (4) at (3,0) {};
\node[] (5) at (4,0) {};
\node[unode] (6) at (5,0) {};
\node[text width=.2cm](33) at (5.1,-0.5){$N_{\alpha}$};
\node[unode] (7) at (7,0) {};
\node[text width=.2cm](34) at (7.1,-0.5){$N_{\alpha+1}$};
\node[text width=.2cm](35) at (5.5,0.3){$M_{\alpha\,\alpha+1}$};
\node[sunode] (8) at (9,0) {};
\node[text width=.2cm] (40) at (9.1,-0.5) {$N_{\alpha+2}$};
\node[fnode] (9) at (9,-2) {};
\node[text width=.2cm] (40) at (9.5,-2) {$M_{\alpha+2}$};
\node[] (10) at (10,0) {};
\node[] (11) at (12,0) {};
\node[afnode] (12) at (7,2) {};
\node[text width=.2cm](36) at (7.5,2){$F$};
\node[fnode] (20) at (0,-2) {};
\node[fnode] (21) at (2,-2) {};
\node[fnode] (22) at (5,-2) {};
\node[text width=.2cm](23) at (0.5,-2){$M_1$};
\node[text width=.2cm](24) at (2.5,-2){$M_2$};
\node[text width=.2cm](25) at (5.5,-2){$M_\alpha$};
\draw[thick] (1) -- (2);
\draw[line width=0.75mm, blue] (2) -- (3);
\node[text width=.2cm](50) at (0.5,0.3){$(\wt{Q}^1,\wt{Q}^2)$};
\node[text width=.2cm](51) at (1,-0.3){$P$};
\draw[thick] (3) -- (4);
\draw[thick,dashed] (4) -- (5);
\draw[thick] (5) -- (6);
\draw[line width=0.75mm] (6) -- (7);
\draw[thick] (7) -- (8);
\draw[thick,blue] (7) -- (12);
\draw[thick] (8) -- (9);
\draw[thick] (8) -- (10);
\draw[thick,dashed] (10) -- (11);
\draw[thick,dashed] (1) -- (100);
\draw[thick, blue] (2) -- (0,1.5);
\draw[thick, blue] (0,1.5) -- (5,1.5);
\draw[thick, blue] (2,1.5) -- (3);
\draw[thick, blue] (5,1.5) -- (6);
\draw[thick] (2) -- (20);
\draw[thick] (3) -- (21);
\draw[thick] (6) -- (22);
\node[text width=.2cm](15) at (0.25,1){$Q^1$};
\node[text width=.2cm](16) at (2.25,1){$Q^2$};
\node[text width=.2cm](17) at (5.25,1){$Q^\alpha$};
\end{tikzpicture}}
\end{center}
\caption{\footnotesize{A generic unitary/special unitary quiver with Abelian hypermultiplets. 
A blue square box with label $F$ represents $F$ Abelian hypermultiplets in the determinant representation. A thin 
blue line connecting multiple unitary gauge nodes is an Abelian hypermultiplet with charges $\{Q^i\}$. A thick blue line with a 
label $P$ denotes a collection of $P$ Abelian hypermultiplets.}}
\label{fig: AbHMgen}
\end{figure}
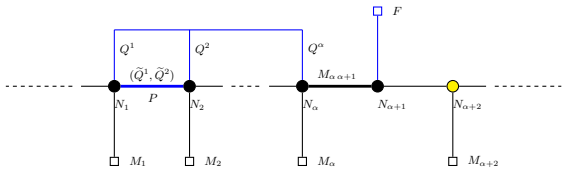

In this paper, we will be interested in a slightly more general theory -- a unitary/special unitary quiver as above with certain additional 
hypermultiplets that transform in powers of the determinant and/or the anti-determinant representations \cite{Dey:2020hfe, Dey:2021rxw} 
of the unitary gauge nodes. We will collectively refer to these matter multiplets as \textit{Abelian Hypermultiplets}. A generic quiver gauge theory of 
this class is given in \figref{fig: AbHMgen}. The simplest quiver gauge theory of this class is a $U(N)$ theory with 
$N_f$ fundamental hypermultiplets and $P$ hypermultiplets in the determinant representation, which we will denote as $\CT^N_{N_f, P}$. 
For $P \geq 1$, these theories are good if $N_f \geq 2N-1$, and bad otherwise.

\noindent \textit{Outline of the paper.} For certain ranges of $N_f$ and $P$, the theory $\CT^N_{N_f, P}$ can be shown to have an IR dual, 
where the duality maps the CB (HB) of one theory to the CB (HB) of the other in the deep IR. Using these dualities one can construct a set of four 
distinct quiver mutations which act locally at appropriate gauge nodes of a quiver having the generic form of \Figref{fig: AbHMgen}.
Any two quivers, which are related by a mutation, flow to the same SCFT in the IR, and are therefore IR dual by construction. 

One can then show that starting from a given theory $\CT$ having the generic form of \Figref{fig: USUgen} (note that it is a special case of the 
quiver in \Figref{fig: AbHMgen}), one can construct a sequence of IR dualities by implementing these quiver mutations. 
The duality sequence leads to $N \geq 2$ distinct quiver gauge theories which flow to the same IR SCFT and are therefore IR dual to each other. 
We refer to this phenomenon as \textit{IR N-ality}. A generic \textit{N-al} theory will be of the form given in \Figref{fig: AbHMgen}. 

Recall that the theory $\CT$ has an emergent IR CB symmetry. 
We show that the set of N-al theories includes at least one theory -- $\CT_{\rm maximal}$ -- for which the rank of the IR
CB symmetry becomes UV-manifest. For $\CT$ being a linear quiver, the complete symmetry algebra itself can be read off from the quiver 
$\CT_{\rm maximal}$. One of the main results of this paper is to give a clear recipe for constructing the quiver $\CT_{\rm maximal}$ 
given $\CT$ and present an illustrative example. \\

\noindent \textit{The IR Dualities of $\CT^N_{N_f, P}$.} We will denote the IR dualities of $\CT^N_{N_f, P}$ as $\CD^N_{N_f, P}$ indicating that there is always 
a $\CT^N_{N_f, P}$ theory on one side. It was shown in \cite{Dey:2022ctw} that there exist three infinite families of such IR dualities, which 
are summarized in \Tabref{Tab: Result-1}. 

\begin{center}
\begin{table}[htbp]
{%
\begin{tabular}{|c|c|c|}
\hline
Duality &Theory & IR dual  \\
\hline \hline 
$\CD^N_{2N+1,1}$
& \scalebox{.7}{\begin{tikzpicture}
\node[unode] (1) at (0,0){};
\node[fnode] (2) at (0,-2){};
\node[afnode] (3) at (3,0){};
\draw[-] (1) -- (2);
\draw[-, thick, blue] (1)-- (3);
\node[text width=0.1cm](10) at (0, 0.5){$N$};
\node[text width=1.5cm](11) at (1, -2){$2N+1$};
\node[text width=1cm](12) at (4, 0){$1$};
\end{tikzpicture}}
&\scalebox{.7}{\begin{tikzpicture}
\node[sunode] (1) at (0,0){};
\node[fnode] (2) at (0,-2){};
\draw[-] (1) -- (2);
\node[text width=1 cm](10) at (0, 0.5){$N+1$};
\node[text width=1.5cm](11) at (1, -2){$2N+1$};
\end{tikzpicture}}\\
\hline
$\CD^N_{2N,P}$
&\scalebox{.6}{\begin{tikzpicture}
\node[unode] (1) at (0,0){};
\node[fnode] (2) at (0,-2){};
\node[afnode] (3) at (3,0){};
\draw[-] (1) -- (2);
\draw[-, thick, blue] (1)-- (3);
\node[text width=0.1cm](10) at (0, 0.5){$N$};
\node[text width=1.5cm](11) at (1, -2){$2N$};
\node[text width=1cm](12) at (4, 0){$P$};
\end{tikzpicture}}  
& \scalebox{.6}{\begin{tikzpicture}
\node[unode] (1) at (0,0){};
\node[fnode] (2) at (0,-2){};
\node[afnode] (3) at (3,0){};
\draw[-] (1) -- (2);
\draw[-, thick, blue] (1)-- (3);
\node[text width=0.1cm](10) at (0, 0.5){$N$};
\node[text width=1.5cm](11) at (1, -2){$2N$};
\node[text width=1cm](12) at (4, 0){$P$};
\end{tikzpicture}} \\
\hline
$\CD^N_{2N-1,P}$
& \scalebox{.6}{\begin{tikzpicture}
\node[unode] (1) at (0,0){};
\node[fnode] (2) at (0,-2){};
\node[afnode] (3) at (3,0){};
\draw[-] (1) -- (2);
\draw[-, thick, blue] (1)-- (3);
\node[text width=0.1cm](10) at (0, 0.5){$N$};
\node[text width=1.5cm](11) at (1, -2){$2N-1$};
\node[text width=1cm](12) at (4, 0){$P$};
\end{tikzpicture}}
 &\scalebox{.6}{\begin{tikzpicture}
\node[unode] (1) at (0,0){};
\node[fnode] (2) at (0,-2){};
\node[unode] (3) at (4,0){};
\node[fnode] (4) at (4,-2){};
\draw[-] (1) -- (2);
\draw[line width=0.75mm, blue] (1)-- (3);
\draw[-] (3)-- (4);
\node[text width=2.3cm](10) at (2.0, 0.2){$(1, -(N-1))$};
\node[text width=2cm](11) at (3.0, -0.3){$P$};
\node[text width=0.1 cm](20) at (0,0.5){$1$};
\node[text width=0.1 cm](21) at (0.5,-2){$1$};
\node[text width=1 cm](22) at (4, 0.5){$N-1$};
\node[text width=1.5 cm](23) at (5, -2){$2N-1$};
\end{tikzpicture}} \\
\hline
\end{tabular}}
\caption{\footnotesize{Summary of the IR dualities for the $\CT^N_{N_f,P}$ theories.}}
\label{Tab: Result-1}
\end{table}
\end{center}

In this notation, the duality $\CD^N_{2N-1,0}$ is the well-known IR duality for an ugly theory \cite{Gaiotto:2008ak} -- it has a $\CT^N_{2N-1,0}$ theory on one side and a $\CT^{N-1}_{2N-1,0}$ theory plus a decoupled twisted hypermultiplet (a $\CT^1_{1,0}$ theory) on the other. The dualities in \Tabref{Tab: Result-1} 
are related to each other as well as the duality $\CD^N_{2N-1,0}$ by various Abelian gauging operations and RG flows triggered 
by large mass parameters, forming a ``duality web"  \cite{Dey:2022ctw}. The dualities can also be checked independently by 
matching supersymmetric observables like the $S^3$ partition function \cite{Kapustin:2010xq} and the supersymmetric index on 
$S^2 \times S^1$ in the Coulomb/Higgs limits \cite{Cremonesi:2013lqa, Razamat:2014pta} --  we refer the reader to Section 3 of \cite{Dey:2022ctw} for details. In the appendix, we summarize the $S^3$ partition function identities for these dualities.

Let us now discuss how the CB symmetry matches across these dualities. For the duality $\CD^N_{2N+1,1}$, one 
has a balanced $SU(N+1)$ gauge theory on one side. This theory has no UV-manifest CB global symmetry, but it does have 
an emergent $\fru(1)$ symmetry. This can be verified, for example, by computing the CB Hilbert Series of the theory. On the other side of the duality, 
this emergent symmetry appears as the UV-manifest $\fru(1)$ topological symmetry of the $U(N)$ gauge group in $\CT^N_{N+1,1}$. 
The duality $\CD^N_{2N,P}$ is the self-duality of the theory $\CT^N_{2N,P}$ which does not have an emergent symmetry. 

For $\CD^N_{2N-1,P}$ with $P \geq 1$, the theory $\CT^N_{2N-1,P}$ has a $\fru(1)$ topological symmetry, and an emergent symmetry 
algebra $\fru(1) \oplus \fru(1)$ for $P >1$ and $\frsu(2) \oplus \fru(1)$ for $P=1$. On the dual side, two $\fru(1)$ factors are manifest in the UV 
as topological symmetries of the $U(1)$ and the $U(N-1)$ gauge groups respectively, thereby matching the rank of the emergent symmetry 
of $\CT^N_{2N-1,P}$. For $P=1$, one can in fact read off the complete IR symmetry from the dual quiver using the result \eref{LQ-IRsymm} for linear quivers. 
Let us think of the dual quiver as being constituted of two linear quivers connected by an Abelian hypermultiplet. 
The $U(1)$ gauge node is balanced and contributes an $\frsu(2)$ factor according to \eref{LQ-IRsymm}, while the $U(N-1)$ gauge node is 
over-balanced and contributes a $\fru(1)$ factor. Therefore, one can visually read off the IR symmetry from the dual quiver as $\frsu(2) \oplus \fru(1)$, 
which is precisely the emergent symmetry of $\CT^N_{2N-1,1}$.

From the above dualities, we learn that a balanced $SU(N)$ gauge node and a $U(N)$ gauge node with balance parameter $e=-1$ plus Abelian hyper(s) 
have emergent CB symmetries, while overbalanced $SU(N)$ nodes and $U(N)$ nodes with $e \geq 0$ do not. 
This will be an important observation for our construction of $\CT_{\rm maximal}$. \\

\noindent \textit{Quiver Mutations and Duality Sequence.}
Given the dualities in \Tabref{Tab: Result-1}, one can construct four distinct quiver mutations which act on the different gauge nodes of a 
quiver gauge theory $\CT$ of the generic form given in \Figref{fig: AbHMgen}. It turns out that for constructing the theory $\CT_{\rm maximal}$, 
it is sufficient to study the sequence of IR dualities generated by only two of the four quiver mutations. We discuss the details of these two 
mutations below, while the remaining two are summarized in the appendix. For more details on these mutations 
and additional examples, we refer the reader to \cite{Dey:2022eko}.

The first mutation, which we will refer to as mutation $I$ and the associated quiver operation as $\CO_I$, involves replacing a balanced $SU$ node by a unitary node of the same rank and a single Abelian hypermultiplet, as shown in \eref{fig: Mutations I}. This mutation is obtained by using the duality $\CD^N_{2N+1,1}$ in the reverse direction.
The Abelian hyper is charged under $U(N_\alpha -1)$ as well as under the unitary gauge nodes connected to it by bifundamental hypers, with  
the charge vector being of the generic form: 
\be \label{charge-1}
\vec Q=(0,\ldots, N_{\alpha _1}, N_{\alpha _2}, -(N_\alpha -1), N_{\alpha _3}, N_{\alpha _4}, \ldots,0),
\ee
where $\{N_{\alpha_i}\}$ denote the ranks of the connected gauge nodes.

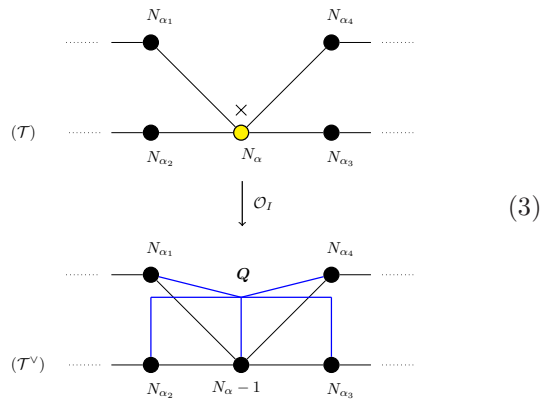
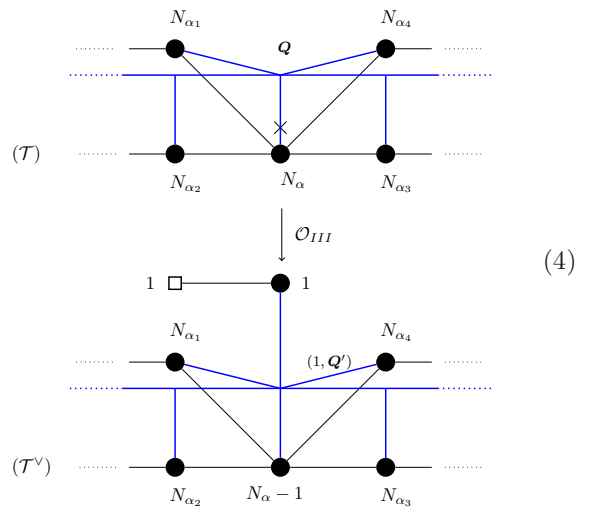
\begin{center}
\be \label{fig: Mutations I}
\begin{tabular}{c}
\scalebox{0.6}{\begin{tikzpicture}
\node[] (1) at (1,0){};
\node[] (100) at (0,0){};
\node[unode] (2) at (2,0){};
\node[sunode] (3) at (4,0){};
\node[unode] (4) at (6,0){};
\node[unode] (51) at (2,2){};
\node[] (52) at (0,2){};
\node[] (53) at (1,2){};
\node[unode] (61) at (6,2){};
\node[] (62) at (7,2){};
\node[] (63) at (8,2){};
\node[] (5) at (7,0){};
\node[] (200) at (8,0){};
\node[cross, red] (6) at (4,0.5){};
\draw[-] (1) -- (2);
\draw[-] (2)-- (3);
\draw[-] (3) -- (4);
\draw[-] (4) --(5);
\draw[-] (3) --(51);
\draw[-] (3) --(61);
\draw[-, dotted] (1) -- (100);
\draw[-,dotted] (5) -- (200);
\draw[-, dotted] (52) -- (53);
\draw[-] (51) -- (53);
\draw[-, dotted] (62) -- (63);
\draw[-] (61) -- (62);
\node[text width=.2cm](11) [below=0.1cm of 2]{$N_{\alpha_2}$};
\node[text width=.2cm](12) at (4.1, -0.5){$N_\alpha$};
\node[text width=.2cm](13) [below=0.1cm of 4]{$N_{\alpha_3}$};
\node[text width=.2cm](15) [above=0.1cm of 51]{$N_{\alpha_1}$};
\node[text width=.2cm](16) [above=0.1cm of 61]{$N_{\alpha_4}$};
\node[text width=.2cm](20) at (-1,0){$(\CT)$};
\end{tikzpicture}} \\
 \scalebox{.6}{\begin{tikzpicture}
\draw[thick, ->] (0,0) -- (0,-1);
\node[text width=0.1cm](29) at (0. 3, -0.5) {$\CO_{I}$};
\node[](30) at (-1.5, -.5) {};
\end{tikzpicture}} \\
\scalebox{0.6}{\begin{tikzpicture}
\node[] (1) at (1,0){};
\node[] (100) at (0,0){};
\node[unode] (2) at (2,0){};
\node[unode] (3) at (4,0){};
\node[unode] (4) at (6,0){};
\node[unode] (51) at (2,2){};
\node[] (52) at (0,2){};
\node[] (53) at (1,2){};
\node[unode] (61) at (6,2){};
\node[] (62) at (7,2){};
\node[] (63) at (8,2){};
\node[] (5) at (7,0){};
\node[] (200) at (8,0){};
\draw[-] (1) -- (2);
\draw[-] (2)-- (3);
\draw[-] (3) -- (4);
\draw[-] (4) --(5);
\draw[-] (3) --(51);
\draw[-] (3) --(61);
\draw[-, dotted] (1) -- (100);
\draw[-,dotted] (5) -- (200);
\draw[-, dotted] (52) -- (53);
\draw[-] (51) -- (53);
\draw[-, dotted] (62) -- (63);
\draw[-] (61) -- (62);
\draw[-, thick, blue] (2)--(2,1.5);
\draw[-, thick, blue] (3)--(4,1.5);
\draw[-, thick, blue] (4)--(6,1.5);
\draw[-, thick, blue] (2,1.5)--(4,1.5);
\draw[-, thick, blue] (4,1.5)--(6,1.5);
\draw[-, thick, blue] (4,1.5)--(51);
\draw[-, thick, blue] (4,1.5)--(61);
\node[text width=.2cm](11) [below=0.1cm of 2]{$N_{\alpha_2}$};
\node[text width=1.5cm](12) at (4.1, -0.5){$N_\alpha -1$};
\node[text width=.2cm](13) [below=0.1cm of 4]{$N_{\alpha_3}$};
\node[text width=.2cm](15) [above=0.1cm of 51]{$N_{\alpha_1}$};
\node[text width=.2cm](16) [above=0.1cm of 61]{$N_{\alpha_4}$};
\node[text width=.2cm](20) at (-1, 0){$(\CT^\vee)$};
\node[text width=.2cm](21) at (4, 2){$\vec Q$};
\end{tikzpicture}}
\end{tabular}
\ee
\end{center}

\be \label{fig: Mutations III}
\begin{tabular}{c}
\scalebox{0.7}{\begin{tikzpicture}
\node[] (1) at (1,0){};
\node[] (100) at (0,0){};
\node[unode] (2) at (2,0){};
\node[unode] (3) at (4,0){};
\node[unode] (4) at (6,0){};
\node[unode] (51) at (2,2){};
\node[] (52) at (0,2){};
\node[] (53) at (1,2){};
\node[unode] (61) at (6,2){};
\node[] (62) at (7,2){};
\node[] (63) at (8,2){};
\node[] (5) at (7,0){};
\node[] (200) at (8,0){};
\node[cross, red] (6) at (4,0.5){};
\draw[-] (1) -- (2);
\draw[-] (2)-- (3);
\draw[-] (3) -- (4);
\draw[-] (4) --(5);
\draw[-] (3) --(51);
\draw[-] (3) --(61);
\draw[-, dotted] (1) -- (100);
\draw[-,dotted] (5) -- (200);
\draw[-, dotted] (52) -- (53);
\draw[-] (51) -- (53);
\draw[-, dotted] (62) -- (63);
\draw[-] (61) -- (62);
\draw[dotted, thick, blue] (0,1.5)--(1,1.5);
\draw[-, thick, blue] (1,1.5)--(2,1.5);
\draw[-, thick, blue] (2)--(2,1.5);
\draw[-, thick, blue] (3)--(4,1.5);
\draw[-, thick, blue] (4)--(6,1.5);
\draw[-, thick, blue] (2,1.5)--(4,1.5);
\draw[-, thick, blue] (4,1.5)--(6,1.5);
\draw[-, thick, blue] (4,1.5)--(51);
\draw[-, thick, blue] (4,1.5)--(61);
\draw[-, thick, blue] (7,1.5)--(6,1.5);
\draw[dotted, thick, blue] (8,1.5)--(7,1.5);
\node[text width=.2cm](11) [below=0.1cm of 2]{$N_{\alpha_2}$};
\node[text width=.2cm](12) at (4.1, -0.5){$N_\alpha$};
\node[text width=.2cm](13) [below=0.1cm of 4]{$N_{\alpha_3}$};
\node[text width=.2cm](15) [above=0.1cm of 51]{$N_{\alpha_1}$};
\node[text width=.2cm](16) [above=0.1cm of 61]{$N_{\alpha_4}$};
\node[text width=0.1 cm](10) at (4, 2){\footnotesize{$\vec Q$}};
\node[text width=.2cm](20) at (-1,0){$(\CT)$};
\end{tikzpicture}} \\
\scalebox{.7}{\begin{tikzpicture}
\draw[->] (0,0) -- (0,-1);
\node[text width=0.1cm](29) at (0.3, -0.5) {$\CO_{III}$};
\node[](30) at (-1.5, -.5) {};
\end{tikzpicture}}\\
 \scalebox{0.7}{\begin{tikzpicture}
\node[] (1) at (1,0){};
\node[] (100) at (0,0){};
\node[unode] (2) at (2,0){};
\node[unode] (3) at (4,0){};
\node[unode] (4) at (6,0){};
\node[unode] (9) at (4,3.5){};
\node[fnode] (10) at (2,3.5){};
\node[unode] (51) at (2,2){};
\node[] (52) at (0,2){};
\node[] (53) at (1,2){};
\node[unode] (61) at (6,2){};
\node[] (62) at (7,2){};
\node[] (63) at (8,2){};
\node[] (5) at (7,0){};
\node[] (200) at (8,0){};
\draw[-] (1) -- (2);
\draw[-] (2)-- (3);
\draw[-] (3) -- (4);
\draw[-] (4) --(5);
\draw[-] (3) --(51);
\draw[-] (3) --(61);
\draw[-] (9) --(10);
\draw[-, dotted] (1) -- (100);
\draw[-,dotted] (5) -- (200);
\draw[-, dotted] (52) -- (53);
\draw[-] (51) -- (53);
\draw[-, dotted] (62) -- (63);
\draw[-] (61) -- (62);
\draw[dotted, thick, blue] (0,1.5)--(1,1.5);
\draw[-, thick, blue] (1,1.5)--(2,1.5);
\draw[-, thick, blue] (2)--(2,1.5);
\draw[-, thick, blue] (3)--(4,1.5);
\draw[-, thick, blue] (4)--(6,1.5);
\draw[-, thick, blue] (2,1.5)--(4,1.5);
\draw[-, thick, blue] (4,1.5)--(6,1.5);
\draw[-, thick, blue] (4,1.5)--(51);
\draw[-, thick, blue] (4,1.5)--(61);
\draw[-, thick, blue] (9)--(4,1.5);
\draw[-, thick, blue] (7,1.5)--(6,1.5);
\draw[dotted, thick, blue] (8,1.5)--(7,1.5);
\node[text width=.2cm](11) [below=0.1cm of 2]{$N_{\alpha_2}$};
\node[text width=1.5cm](12) at (4.1, -0.5){$N_\alpha -1$};
\node[text width=.2cm](13) [below=0.1cm of 4]{$N_{\alpha_3}$};
\node[text width=.2cm](24) [right=0.1cm of 9]{$1$};
\node[text width=.2cm](25) [left=0.1cm of 10]{$1$};
\node[text width=.2cm](15) [above=0.1cm of 51]{$N_{\alpha_1}$};
\node[text width=.2cm](16) [above=0.1cm of 61]{$N_{\alpha_4}$};
\node[text width=2 cm](30) at (5.5, 2){\footnotesize{$(1,\vec Q')$}};
\node[text width=.2cm](20) at (-1, 0){$(\CT^\vee)$};
\end{tikzpicture}}
\end{tabular}
\ee

The three remaining mutations act on $U(N_\alpha)$ gauge nodes with Abelian hypermultiplets, and 
correspond to following values of the balance parameter $e_\alpha=1,0,-1$. Mutation $I'$ and Mutation $II$ 
(with associated quiver operations $\CO_{I'}$ and $\CO_{II}$ respectively) act on gauge nodes with balance parameters $e_\alpha = 1$ and $e_\alpha = 0$ 
respectively, and are not relevant for the construction of $\CT_{\rm maximal}$ (we will explain why momentarily). We discuss these mutations in the appendix. 

Mutation $III$ (quiver operation $\CO_{III}$) corresponds to the case $e_\alpha =-1$, and is obtained by using the duality $\CD^N_{2N-1,P}$. 
The mutation splits the $U(N_\alpha)$ gauge node into a $U(N_\alpha-1)$ node and a $U(1)$ node with the latter node having a single fundamental hyper, 
as shown in \eref{fig: Mutations III} for the $P=1$ case. The $P$ Abelian hypers in $\CT$ of charges $\{ \vec Q^l\}_{l=1,\ldots,P}$ are mapped to another $P$ Abelian hypers in 
$\CT^\vee$. The latter Abelian hypers all have charge 1 under the new $U(1)$ node and have charges $\{\vec Q'^l\}_{l=1,\ldots,P}$ under the remaining gauge nodes. 
For a generic $ \vec Q^l= (Q^l_1, \ldots, Q^l_{\alpha_1}, Q^l_{\alpha_2}, N_\alpha, Q^l_{\alpha_3}, Q^l_{\alpha_4}, \ldots, Q^l_L)$, the charge vector $\vec Q'^l$ is given as 
\begin{align} \label{charge-3}
\vec Q'^l= &(Q^l_1, \ldots, Q^l_{\alpha_1} + N_{\alpha_1}, Q^l_{\alpha_2} + N_{\alpha_2},  -(N_\alpha-1), \nn \\
 & Q^l_{\alpha_3} + N_{\alpha_3} , Q^l_{\alpha_4} + N_{\alpha_4}, \ldots, Q^l_L),
\end{align}
where $\{N_{\alpha_i}\}$ denote the ranks of the nodes connected to $U(N_\alpha)$ by bifundamental hypers. Note that only the charges associated with the nodes connected to $U(N_\alpha)$ with bifundamental hypers get transformed under the mutation. The mutations can be realized in terms of supersymmetric observables -- we will discuss the $S^3$ partition function realization in the appendix.

Let us now consider a theory $\CT$ in the class of theories of \figref{fig: USUgen}.
As we saw above, a balanced $SU$ node is associated with a $\fru(1)$ emergent symmetry.  
In the presence of balanced unitary nodes connected to this balanced $SU$ node, the CB symmetry may be further enhanced. 
As before, the emergent symmetry can be verified using the CB limit of the index. Given the quiver mutations discussed above, 
the duality sequence leading to the theory $\CT_{\rm maximal}$ can be obtained in the following fashion. 

One begins by first implementing mutation $I$ at every balanced $SU$ node in $\CT$. Other $SU$ nodes which were 
overbalanced in $\CT$ might be rendered balanced as a result, in which case we implement mutation $I$ sequentially until we 
have a theory that contains no balanced $SU$ nodes. In the next step, one implements mutation $III$ at every gauge node 
that admits it. In doing so, one will generically alter the balance of both unitary and special unitary nodes in the quiver, thereby creating 
new nodes where mutation $III$ or mutation $I$ can be implemented.  The duality sequence finally terminates at a quiver for which none of the 
gauge nodes admit either mutation $I$ or mutation $III$. This quiver therefore consists of overbalanced special unitary nodes and unitary nodes of balance parameters 
$e \geq 0$ with or without Abelian hypers. Since neither type of gauge nodes leads to emergent CB symmetry, one expects that the UV-manifest rank 
should match the rank of the IR symmetry of the quiver. The theory is therefore a candidate for $\CT_{\rm maximal}$.

The quiver operations $\CO_I$ and $\CO_{III}$ increase the number of $\fru(1)$ topological symmetries by 1, $\CO_{I'}$ decreases it by 1, 
and $\CO_{II}$ keeps it invariant. This is the reason why one can ignore $\CO_{I'}$ and $\CO_{II}$ if one is interested in finding a single 
candidate for $\CT_{\rm maximal}$. However, the complete duality sequence must include these mutations as well. In particular, 
there may be multiple candidates for $\CT_{\rm maximal}$ which are related by $\CO_{II}$. In addition, the operation $\CO_{I'}$ arises 
in the closure relations of  $\CO_I$ and $\CO_{III}$, as we discuss in the appendix.\\

\noindent \textit{An Illustrative Example.}
In this section, we will construct the duality sequence for a linear quiver with unitary/special unitary gauge nodes and 
determine $\CT_{\rm maximal}$ explicitly. We will show that it is possible to read off the emergent CB symmetry algebra 
$\frg^{\rm IR}_{\rm C}$ from the quiver representation of $\CT_{\rm maximal}$. 
Consider a three-node quiver $\CT$ with a single $SU$ node of the following form: 
\begin{center}
\scalebox{0.65}{\begin{tikzpicture}
\node[fnode] (1) {};
\node[unode] (2) [right=.75cm  of 1]{};
\node[sunode] (3) [right=.75cm of 2]{};
\node[unode] (4) [right=0.75 cm of 3]{};
\node[fnode] (5) [right=0.75 cm of 4]{};
\draw[-] (1) -- (2);
\draw[-] (2)-- (3);
\draw[-] (3) -- (4);
\draw[-] (4) -- (5);
\node[text width=.1cm](10) [below=0.1 cm of 1]{$M_1$};
\node[text width=.2cm](11) [below=0.1cm of 2]{$N_1$};
\node[text width=.1cm](12) [below=0.1cm of 3]{$N$};
\node[text width=.1cm](13) [below=0.1cm of 4]{$N_2$};
\node[text width=.1cm](14) [below=0.1cm of 5]{$M_2$};
\node[text width=.2cm](20) [left= 1 cm of 1]{$(\CT):$};
\end{tikzpicture}}
\end{center}

We will focus on the case where the central $SU(N)$ gauge node as well as the two unitary nodes are balanced i.e. $N_1 + N_2=2N-1$, $M_1 + N =2N_1$ and $M_2 + N =2N_2$. The theory has an emergent symmetry $\frg^{\rm IR}_{\rm C}(\CT)= \frsu(2) \oplus \frsu(2) \oplus \frsu(4) \oplus \fru(1)$. In particular, the rank of the emergent symmetry $\text{rk}(\frg^{\rm IR}_{\rm C}(\CT)) = 6$ is manifestly different from the rank of the UV symmetry $\text{rk}(\frg^{\rm UV}_{\rm C}(\CT)) = 2$.

The first step for constructing the duality sequence is to implement mutation $I$ on the balanced $SU(N)$ node following \eref{fig: Mutations I}: 
\begin{center}
\begin{tabular}{ccc}
\scalebox{0.6}{\begin{tikzpicture}
\node[fnode] (1) at (2,-1.5){};
\node[unode] (2) at (2,0){};
\node[sunode] (3) at (4,0){};
\node[unode] (4) at (6,0){};
\node[fnode] (5) at (6,-1.5){};
\node[cross, red] (6) at (4,0.5){};
\draw[-] (1) -- (2);
\draw[-] (2)-- (3);
\draw[-] (3) -- (4);
\draw[-] (4) --(5);
\node[text width=.1cm](10) [right=0.1 cm of 1]{$M_1$};
\node[text width=.2cm](11) [below=0.1cm of 2]{$N_1$};
\node[text width=.1cm](12) [below=0.1cm of 3]{$N$};
\node[text width=.1cm](13) [below=0.1cm of 4]{$N_2$};
\node[text width=.1cm](14) [right=0.1cm of 5]{$M_2$};
\node[text width=.2cm](20) at (4,-2){$(\CT)$};
\end{tikzpicture}}
& \scalebox{.6}{\begin{tikzpicture}
\draw[->] (4,0) -- (6, 0);
\node[text width=0.1cm](29) at (5, 0.2) {$\CO_I$};
\node[](30) at (5, -2.1) {};
\end{tikzpicture}}
&\scalebox{0.6}{\begin{tikzpicture}
\node[fnode] (1) at (2,-1.5){};
\node[unode] (2) at (2,0){};
\node[unode] (3) at (4,0){};
\node[unode] (4) at (6,0){};
\node[fnode] (5) at (6,-1.5){};
\draw[-] (1) -- (2);
\draw[-] (2)-- (3);
\draw[-] (3) -- (4);
\draw[-] (4) --(5);
\draw[-, thick, blue] (2)--(2,1.5);
\draw[-, thick, blue] (3)--(4,1.5);
\draw[-, thick, blue] (4)--(6,1.5);
\draw[-, thick, blue] (2,1.5)--(4,1.5);
\draw[-, thick, blue] (4,1.5)--(6,1.5);
\node[text width=3 cm](10) at (4, 2){\footnotesize{$(N_1, -(N-1), N_2)$}};
\node[text width=.1cm](10) [right=0.1 cm of 1]{$M_1$};
\node[text width=.2cm](11) [below=0.1cm of 2]{$N_1$};
\node[text width= 1cm](12) [below=0.1cm of 3]{$N-1$};
\node[text width=.1cm](13) [below=0.1cm of 4]{$N_2$};
\node[text width=.1cm](14) [right=0.1cm of 5]{$M_2$};
\node[text width=.2cm](20) at (4,-2){$(\CT^\vee_1)$};
\end{tikzpicture}}
\end{tabular}
\end{center}

The above mutation increases the UV-manifest rank by 1, since $\text{rk}(\frg^{\rm UV}_{\rm C}(\CT^\vee_1)) = 3$, 
as can be seen from the quiver $\CT^\vee_1$. The balance of the first and the third gauge nodes (from the left) are $e_1=e_3=-1$, and therefore 
one can implement the mutation $\CO_{III}$ at each of these nodes. In the second step, we implement mutation $III$ on the leftmost node 
following \eref{fig: Mutations III} which leads to the quiver $\CT^\vee_2$:

\begin{center}
\begin{tabular}{ccc}
\scalebox{0.6}{\begin{tikzpicture}
\node[fnode] (1) at (2,-1.5){};
\node[unode] (2) at (2,0){};
\node[unode] (3) at (4,0){};
\node[unode] (4) at (6,0){};
\node[fnode] (5) at (6,-1.5){};
\node[cross, red] (6) at (2,0.5){};
\draw[-] (1) -- (2);
\draw[-] (2)-- (3);
\draw[-] (3) -- (4);
\draw[-] (4) --(5);
\draw[-, thick, blue] (2)--(2,1.5);
\draw[-, thick, blue] (3)--(4,1.5);
\draw[-, thick, blue] (4)--(6,1.5);
\draw[-, thick, blue] (2,1.5)--(4,1.5);
\draw[-, thick, blue] (4,1.5)--(6,1.5);
\node[text width=3 cm](10) at (4, 2){\footnotesize{$(N_1, -(N-1), N_2)$}};
\node[text width=.1cm](10) [right=0.1 cm of 1]{$M_1$};
\node[text width=.2cm](11) [below=0.1cm of 2]{$N_1$};
\node[text width= 1cm](12) [below=0.1cm of 3]{$N-1$};
\node[text width=.1cm](13) [below=0.1cm of 4]{$N_2$};
\node[text width=.1cm](14) [right=0.1cm of 5]{$M_2$};
\node[text width=.2cm](20) at (4,-2){$(\CT^\vee_1)$};
\end{tikzpicture}}
& \scalebox{.6}{\begin{tikzpicture}
\draw[->] (4,0) -- (6, 0);
\node[text width=0.2cm](29) at (5, 0.2) {$\CO_{III}$};
\node[](30) at (5, -2.1) {};
\end{tikzpicture}}
&\scalebox{0.6}{\begin{tikzpicture}
\node[fnode] (1) at (2,-1.5){};
\node[unode] (2) at (2,0){};
\node[unode] (3) at (4,0){};
\node[unode] (4) at (6,0){};
\node[fnode] (5) at (6,-1.5){};
\node[unode] (6) at (2,3){};
\node[fnode] (7) at (4,3){};
\draw[-] (1) -- (2);
\draw[-] (2)-- (3);
\draw[-] (3) -- (4);
\draw[-] (4) --(5);
\draw[-] (6) --(7);
\draw[-, thick, blue] (2)--(6);
\draw[-, thick, blue] (4)--(6,1.5);
\draw[-, thick, blue] (2,1.5)--(4,1.5);
\draw[-, thick, blue] (4,1.5)--(6,1.5);
\node[text width=3 cm](40) at (4, 2){\footnotesize{$(1, -(N_1-1), N_2)$}};
\node[text width=.1cm](20) [right=0.2 cm of 1]{$M_1$};
\node[text width=1 cm](21) [below=0.1cm of 2]{$N_1-1$};
\node[text width=1 cm](22) [below=0.1cm of 3]{$N-1$};
\node[text width=0.1 cm](23) [below=0.1cm of 4]{$N_2$};
\node[text width=.1cm](24) [right=0.1cm of 5]{$M_2$};
\node[text width=.1cm](25) [left=0.1cm of 6]{1};
\node[text width=.1cm](26) [right=0.1 cm of 7]{1};
\node[text width=.2cm](20) at (4,-2){$(\CT^\vee_2)$};
\end{tikzpicture}}
\end{tabular}
\end{center}

This is followed by the mutation on the rightmost gauge node which leads to the quiver $(\CT^\vee_3)$: 
\begin{center}
\begin{tabular}{ccc}
\scalebox{0.6}{\begin{tikzpicture}
\node[fnode] (1) at (2,-1.5){};
\node[unode] (2) at (2,0){};
\node[unode] (3) at (4,0){};
\node[unode] (4) at (6,0){};
\node[fnode] (5) at (6,-1.5){};
\node[unode] (6) at (2,3){};
\node[fnode] (7) at (4,3){};
\node[cross, red] (8) at (6,0.5){};
\draw[-] (1) -- (2);
\draw[-] (2)-- (3);
\draw[-] (3) -- (4);
\draw[-] (4) --(5);
\draw[-] (6) --(7);
\draw[-, thick, blue] (2)--(6);
\draw[-, thick, blue] (4)--(6,1.5);
\draw[-, thick, blue] (2,1.5)--(4,1.5);
\draw[-, thick, blue] (4,1.5)--(6,1.5);
\node[text width=3 cm](40) at (4, 2){\footnotesize{$(1, -(N_1-1), N_2)$}};
\node[text width=.1cm](20) [right=0.2 cm of 1]{$M_1$};
\node[text width=1 cm](21) [below=0.1cm of 2]{$N_1-1$};
\node[text width=1 cm](22) [below=0.1cm of 3]{$N-1$};
\node[text width=0.1 cm](23) [below=0.1cm of 4]{$N_2$};
\node[text width=.1cm](24) [right=0.1cm of 5]{$M_2$};
\node[text width=.1cm](25) [left=0.1cm of 6]{1};
\node[text width=.1cm](26) [right=0.1 cm of 7]{1};
\node[text width=.2cm](20) at (4,-2){$(\CT^\vee_2)$};
\end{tikzpicture}}
& \scalebox{.6}{\begin{tikzpicture}
\draw[->] (4,0) -- (6, 0);
\node[text width=0.2cm](29) at (5, 0.2) {$\CO_{III}$};
\node[](30) at (5, -2.1) {};
\end{tikzpicture}}
&\scalebox{0.6}{\begin{tikzpicture}
\node[fnode] (1) at (2,-1.5){};
\node[unode] (2) at (2,0){};
\node[unode] (3) at (4,0){};
\node[unode] (4) at (6,0){};
\node[fnode] (5) at (6,-1.5){};
\node[unode] (6) at (2,3){};
\node[fnode] (7) at (1,3){};
\node[unode] (8) at (6,3){};
\node[fnode] (9) at (7,3){};
\draw[-] (1) -- (2);
\draw[-] (2)-- (3);
\draw[-] (3) -- (4);
\draw[-] (4) --(5);
\draw[-] (6) --(7);
\draw[-] (8) --(9);
\draw[-, thick, blue] (6)--(4,1.5);
\draw[-, thick, blue] (3)--(4,1.5);
\draw[-, thick, blue] (8)--(4,1.5);
\draw[-, thick, blue] (2)--(2,1.5);
\draw[-, thick, blue] (4)--(6,1.5);
\draw[-, thick, blue] (2,1.5)--(4,1.5);
\draw[-, thick, blue] (4,1.5)--(6,1.5);
\node[text width=4.5cm](19) at (4.1, 2){\footnotesize{$(1,1, -N_1+1,N-1,-N_2+1)$}};
\node[text width=.2cm](12) at (4,-2){$(\CT^\vee_3)$};
\node[text width=.1cm](20) [right=0.1 cm of 1]{$M_1$};
\node[text width=1 cm](21) [below=0.1cm of 2]{$N_1-1$};
\node[text width=1 cm](22) [below=0.1cm of 3]{$N-1$};
\node[text width=1 cm](23) [below=0.1cm of 4]{$N_2-1$};
\node[text width=.1cm](24) [right=0.1cm of 5]{$M_2$};
\node[text width=.1cm](25) [right=0.1cm of 6]{1};
\node[text width=.1cm](26) [left=0.1 cm of 7]{1};
\node[text width=.1cm](27) [left=0.5 cm of 8]{1};
\node[text width=.1cm](28) [right=0.1 cm of 9]{1};
\end{tikzpicture}}
\end{tabular}
\end{center}

Note that at each step, starting from $\CT^\vee_1$ to $\CT^\vee_3$, the UV-manifest rank of the symmetry increases by 1, 
due to the addition of a single $U(1)$ gauge node. In the quiver $\CT^\vee_3$, the central gauge node has balance $e_2=-1$, and 
one can implement yet another $\CO_{III}$ mutation:

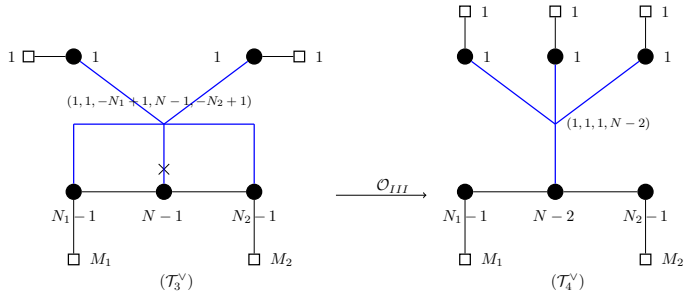
\begin{center}
\begin{tabular}{ccc}
\scalebox{0.6}{\begin{tikzpicture}
\node[fnode] (1) at (2,-1.5){};
\node[unode] (2) at (2,0){};
\node[unode] (3) at (4,0){};
\node[unode] (4) at (6,0){};
\node[fnode] (5) at (6,-1.5){};
\node[unode] (6) at (2,3){};
\node[fnode] (7) at (1,3){};
\node[unode] (8) at (6,3){};
\node[fnode] (9) at (7,3){};
\node[cross, red] (10) at (4,0.5){};
\draw[-] (1) -- (2);
\draw[-] (2)-- (3);
\draw[-] (3) -- (4);
\draw[-] (4) --(5);
\draw[-] (6) --(7);
\draw[-] (8) --(9);
\draw[-, thick, blue] (6)--(4,1.5);
\draw[-, thick, blue] (3)--(4,1.5);
\draw[-, thick, blue] (8)--(4,1.5);
\draw[-, thick, blue] (2)--(2,1.5);
\draw[-, thick, blue] (4)--(6,1.5);
\draw[-, thick, blue] (2,1.5)--(4,1.5);
\draw[-, thick, blue] (4,1.5)--(6,1.5);
\node[text width=4.5cm](19) at (4.1, 2){\footnotesize{$(1,1, -N_1+1,N-1,-N_2+1)$}};
\node[text width=.2cm](12) at (4,-2){$(\CT^\vee_3)$};
\node[text width=.1cm](20) [right=0.1 cm of 1]{$M_1$};
\node[text width=1 cm](21) [below=0.1cm of 2]{$N_1-1$};
\node[text width=1 cm](22) [below=0.1cm of 3]{$N-1$};
\node[text width=1 cm](23) [below=0.1cm of 4]{$N_2-1$};
\node[text width=.1cm](24) [right=0.1cm of 5]{$M_2$};
\node[text width=.1cm](25) [right=0.1cm of 6]{1};
\node[text width=.1cm](26) [left=0.1 cm of 7]{1};
\node[text width=.1cm](27) [left=0.5 cm of 8]{1};
\node[text width=.1cm](28) [right=0.1 cm of 9]{1};
\end{tikzpicture}}
& \scalebox{.6}{\begin{tikzpicture}
\draw[->] (4,0) -- (6, 0);
\node[text width=0.2cm](29) at (5, 0.2) {$\CO_{III}$};
\node[](30) at (5, -2.1) {};
\end{tikzpicture}}
&\scalebox{0.6}{\begin{tikzpicture}
\node[fnode] (1) at (2,-1.5){};
\node[unode] (2) at (2,0){};
\node[unode] (3) at (4,0){};
\node[unode] (4) at (6,0){};
\node[fnode] (5) at (6, -1.5){};
\node[unode] (6) at (2,3){};
\node[fnode] (7) at (2,4){};
\node[unode] (8) at (6,3){};
\node[fnode] (9) at (6,4){};
\node[unode] (22) at (4,3){};
\node[fnode] (23) at (4,4){};
\draw[-] (1) -- (2);
\draw[-] (2)-- (3);
\draw[-] (3) -- (4);
\draw[-] (4) --(5);
\draw[-] (6) --(7);
\draw[-] (8) --(9);
\draw[-] (22) --(23);
\draw[-, thick, blue] (6)--(4,1.5);
\draw[-, thick, blue] (3)--(4,1.5);
\draw[-, thick, blue] (8)--(4,1.5);
\draw[-, thick, blue] (22)--(4,1.5);
\node[text width=4.5cm](19) at (6.5, 1.5){\footnotesize{$(1,1, 1, N-2)$}};
\node[text width=.1cm](40) [right=0.1 cm of 1]{$M_1$};
\node[text width=1 cm](41) [below=0.1cm of 2]{$N_1-1$};
\node[text width=1 cm](42) [below=0.1cm of 3]{$N-2$};
\node[text width=1 cm](43) [below=0.1cm of 4]{$N_2-1$};
\node[text width=.1cm](44) [right=0.1cm of 5]{$M_2$};
\node[text width=.1cm](45) [right=0.1cm of 6]{1};
\node[text width=.1cm](46) [right=0.1 cm of 7]{1};
\node[text width=.1cm](47) [right=0.1 cm of 8]{1};
\node[text width=.1cm](48) [right=0.1 cm of 9]{1};
\node[text width=.1cm](49) [right=0.1 cm of 22]{1};
\node[text width=.1cm](50) [right=0.1 cm of 23]{1};
\node[text width=.2cm](20) at (4,-2){$(\CT^\vee_4)$};
\end{tikzpicture}}
\end{tabular}
\end{center}

The first and the third gauge node (from the left) in $\CT^\vee_4$ are balanced, while the central node is overbalanced i.e. $e_2=1$. 
This implies that one cannot implement another mutation ${III}$. Since there are no $SU$ nodes left, one cannot 
implement a mutation $I$ either. Therefore, following the logic described in the previous section, 
we have 
\be \CT_{\rm maximal} =: \CT^\vee_4. 
\ee 
It is convenient to rewrite the quiver after a simple field redefinition in the following form:
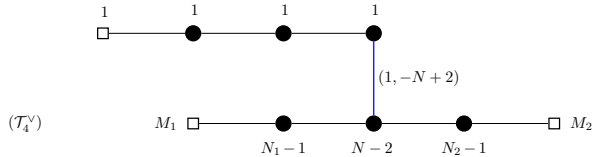
\begin{center}
\scalebox{0.6}{\begin{tikzpicture}
\node[fnode] (1) at (0,0){};
\node[unode] (2) at (2,0){};
\node[unode] (3) at (4,0){};
\node[unode] (4) at (6,0){};
\node[fnode] (5) at (8,0){};
\node[unode] (6) at (4,2){};
\node[unode] (7) at (2,2){};
\node[unode] (8) at (0,2){};
\node[fnode] (9) at (-2,2){};
\draw[-] (1) -- (2);
\draw[-] (2)-- (3);
\draw[-] (3) -- (4);
\draw[-] (4) --(5);
\draw[-] (6) --(7);
\draw[-] (8) --(7);
\draw[-] (8) --(9);
\draw[thick, blue] (6)-- (3);
\node[text width=.1cm](40) [left=0.5 cm of 1]{$M_1$};
\node[text width=1 cm](41) [below=0.1cm of 2]{$N_1-1$};
\node[text width=1 cm](42) [below=0.1cm of 3]{$N-2$};
\node[text width=1 cm](43) [below=0.1cm of 4]{$N_2-1$};
\node[text width=.1cm](44) [right=0.1cm of 5]{$M_2$};
\node[text width=.1cm](45) [above=0.1cm of 6]{1};
\node[text width=.1cm](46) [above=0.1 cm of 7]{1};
\node[text width=.1cm](47) [above=0.1 cm of 8]{1};
\node[text width=.1cm](48) [above=0.1 cm of 9]{1};
\node[text width=2cm](21) at (5.1, 1){$(1,-N+2)$};
\node[text width=.2 cm](20) at (-4,0){$(\CT^\vee_{4})$};
\end{tikzpicture}}
\end{center}

For the quiver $\CT^\vee_4$, the UV-manifest rank can be read off as $\text{rk}(\frg^{\rm UV}_{\rm C}(\CT^\vee_4)) = 6$, which precisely matches the rank of the IR symmetry of $\CT$. Let us now show how one can read off the symmetry algebra $\frg^{\rm IR}_{\rm C}$ itself using our intuition from linear quivers with unitary gauge groups. 

Firstly, note that the quiver $\CT^\vee_4$ is built out of two linear subquivers with unitary gauge groups connected by a single Abelian hyper that 
is charged under a single node in each subquiver. The first subquiver -- a chain of three balanced $U(1)$ nodes -- contributes a factor $\frsu(4)$ 
to the IR symmetry, following \eref{LQ-IRsymm}. In the second subquiver, the balanced nodes $U(N_1-1)$ and $U(N_2-1)$ are expected to 
contribute an $\frsu(2)$ factor each, while the overbalanced central node (connected to the Abelian hyper) gives a $\fru(1)$ factor. Therefore, one 
reads off the IR symmetry of $\CT^\vee_4$ as $\frg^{\rm IR}_{\rm C}(\CT^\vee_4)=\frsu(4) \oplus \frsu(2) \oplus \frsu(2) \oplus \fru(1)$, which precisely 
matches the IR symmetry algebra of $\CT$.\\

\noindent \textit{Conclusion and Outlook.}
A unitary-special unitary quiver gauge theory $\CT$ of generic shape with at least a single balanced $SU$ node 
admits a sequence of IR dualities. This duality sequence can be generated by step-wise implementing four distinct quiver mutations locally at different 
gauge nodes, starting with a balanced $SU$ node. These quiver mutations are in turn built out of IR dualities of $U(N)$ gauge theories 
with $N_f$ hypers in the fundamental representation and $P$ hypers in the determinant representation, for certain ranges of 
$N_f$ and $P$. 

The theory $\CT$ has an emergent CB symmetry characterized by the presence of hidden FI parameters which implies that 
the rank of the IR symmetry is greater than UV-manifest rank. The sequence of dualities provides a neat way to study the 
emergent CB symmetry of $\CT$. We have shown that duality sequence produces at least one theory $\CT_{\rm maximal}$ for which 
the correct rank of the IR symmetry becomes manifest from the quiver description. For a subclass of theories, one may even be 
able to read off the correct symmetry algebra. Using a simple example, we demonstrated that this is indeed the case when $\CT$ is a 
linear quiver.

Our formalism gives the first systematic way to study the emergent CB symmetry (and therefore hidden FI parameters) in 3d $\CN=4$ 
theories which do not have a realization in String Theory (like the Hanany-Witten \cite{Hanany:1996ie} description or a description in terms of 
magnetic quivers \cite{Bourget:2021jwo}). It also leads to an extremely efficient algorithm for generating the 3d mirrors of unitary-special unitary quivers 
with generic shape, which will be presented in a paper to appear soon. Analogous to 3d mirror symmetry, various aspects of these IR dualities 
-- for example, the duality maps for BPS local operators and line defects -- should furnish interesting physics and deserve detailed investigation.
Finally, one expects to find novel non-supersymmetric dualities as one subjects these N-al theories to soft supersymmetry-breaking, in a 
fashion similar to \cite{Kachru:2016rui}. Some of these topics will be addressed in future work. \\

\noindent \textbf{Acknowledgments} The author would like to thank Amihay Hanany and Zohar Komargodski for discussion on related issues, 
and Vivek Saxena for comments on the draft. The author would like to thank the organizers of the program ``Hyperkahler quotients, singularities, and quivers" 
at the Simons Center for Geometry and Physics where results connected to this work were presented. The author
acknowledges the hospitality of the Simons Summer Workshop 2023 during the completion of this work. 
The author is supported in part at the Johns Hopkins University by the NSF grant PHY-2112699.

\bibliographystyle{apsrev4-2}
\bibliography{cpn1-1}

\onecolumngrid

\appendix

\section{The partition function identities}\label{app: S3pf}

In this appendix, we summarize the round three-sphere partition function identities that realize the 
IR dualities listed in \Tabref{Tab: Result-1} and the quiver mutations. We will denote the real masses associated with the fundamental hypers 
and the Abelian hyper in the $\CT^N_{N_f, P}$ theory as $\vec m$ and $\vec{m}_{\rm ab}$ respectively, and the 
single FI parameter as $\eta$. 

Let us begin with the duality $\CD^N_{2N+1,1}$. The corresponding identity is given as:

\be \label{Id-1}
Z^{\CT^{N}_{2N+1,1}}(\vec m, m_{\rm ab}=\tr \vec m, \eta=0) = Z^{\CT^{SU(N+1)}_{2N +1}}(\vec m).
\ee
The $2N+1$ independent mass parameters live in the Cartan subalgebra of the HB global symmetry algebra $\frsu(2N+1) \oplus \fru(1)$.
The equality of the partition functions holds only after the FI parameter of the $U(N)$ vector multiplet is set to zero, which  
is expected since the $SU(N+1)$ theory does not have a UV-manifest $\fru(1)$ topological symmetry for which one can 
turn on an Fl parameter.

Next, for the self-duality $\CD^N_{2N,1}$, the corresponding identity is given as:

\be \label{Id-2}
Z^{\CT^N_{2N,1}}(\vec m, {m}_{\rm ab}, \eta) = e^{2\pi i \eta\,\tr \vec{m}} \cdot Z^{\CT^N_{2N,1}}(\vec m, \tr\vec{m}- {m}_{\rm ab}, -\eta).
\ee
Although the independent masses in this case live in the Cartan subalgebra of the HB global symmetry algebra $\frsu(2N) \oplus \fru(1)$, 
it is convenient to write the identity in the above form (i.e. with a single redundant parameter) for deriving the quiver mutations. The 
extension of the identity to the $P >1$ case is straightforward. 

Finally, consider the duality $\CD^N_{2N-1,1}$, for which the identity assumes the form: 
\be \label{Id-3}
Z^{\CT^N_{2N-1,1}}(\vec m, {m}_{\rm ab}, \eta) = Z^{\CT^\vee}(m^\vee_{(1)},\vec m^\vee_{(2)}, m^\vee_{\rm{ab}};\eta, -\eta),
\ee
where $m^\vee_{(1)},\vec m^\vee_{(2)}, m^\vee_{\rm{ab}}$ are the masses for the $U(1)$ gauge node, the $U(N-1)$ gauge 
node, and the Abelian hyper respectively. In terms of the mass parameters of the theory $\CT^N_{2N-1,1}$, these are given 
as:

\be
m^\vee_{(1)} = \tr \vec m, \quad  \vec m^\vee_{(2)} = \vec m, \quad m^\vee_{\rm{ab}}= m_{\rm{ab}}. 
\ee

Similar to the self-duality $\CD^N_{2N,1}$, we have written the identity with a single additional mass parameter. The independent 
mass parameters are in the Cartan subalgebra of $\frsu(2N-1) \oplus \fru(1)$. The extension to $P > 1$ is again straightforward. 

The quiver mutations discussed in the main text of the paper and the appendix can be realized in terms of the sphere partition 
function by using the above identities locally at a gauge node of a quiver gauge theory. We will briefly describe the cases of mutation $I$ 
and mutation $III$ here, and refer the reader to \cite{Dey:2022eko} for a more extensive discussion. Mutation $I$ is implemented by using the 
identity \eref{Id-1} locally for the $SU(N_\alpha)$ gauge node in the quiver $\CT$ (see the figure in \eref{fig: Mutations I}):
\begin{align}\label{Id-1-main}
Z^{(\CT)}=  \int \, [d \vec s_\alpha] \, \frac{\delta(\tr \vec s_\alpha)\, Z_{\rm vec}(\vec s_\alpha)\, \Big[ \ldots \Big]}{\prod_{\alpha_i}\,\prod_{j,i}\,\ch{(s^j_\alpha - \s^i_{\alpha_i})} } 
& =  \int \, [d \vec \s_\alpha] \, \frac{Z_{\rm vec}(\vec \s_\alpha)\, \Big[ \ldots \Big]}{ \prod_{\alpha_i}\,\prod_{j,i}\,\ch{(\s^j_\alpha - \s^i_{\alpha_i})}\,\ch{(-\tr \vec \s_\alpha 
+\sum_ {\alpha_i}\,\tr \vec \s_{\alpha_i} )}  } \nn \\
&= Z^{(\CT^\vee)}(\eta^\vee_\alpha=0, \ldots), 
\end{align}
where $\Big[ \ldots \Big]$ denotes the terms in the partition function independent of the $SU(N_\alpha)$ node 
and the $U(N_\alpha -1)$ node respectively,  and $\eta^\vee_\alpha$ is the FI parameter of the $U(N_\alpha -1)$ gauge node in $\CT^\vee$. 
The charges for the Abelian hypermultiplet in $\CT^\vee$ can be simply read off from the partition function and seen to agree with \eref{charge-1}.

Mutation $III$ can be realized by implementing the identity \eref{Id-3} locally 
for the $U(N_\alpha)$ gauge node in the quiver $\CT$. For the $P=1$ case, we obtain (see the figure in \eref{fig: Mutations III}):
\begin{align}\label{Id-3-main}
Z^{(\CT)}&=  \int \, [d \vec s_\alpha] \, \frac{ e^{2\pi i \eta_\alpha\, \tr \vec s_\alpha}\, Z_{\rm vec}(\vec s_\alpha)\, \Big[ \ldots \Big]}{ \prod_{\alpha_i}\,\prod_{j,i}\,\ch{(s^j_\alpha - \s^i_{\alpha_i})}\,\ch{(\tr \vec s_\alpha + \sum_a \frac{Q_a}{N_a}\, \tr \vec\s_a)}} \nn \\
&=  \int \, d\s' \, [d \vec \s_\alpha] \, \frac{e^{2\pi i \eta_\alpha\, (\s' -\tr \vec \s_\alpha)}\, Z_{\rm vec}(\vec \s_\alpha)\, \Big[ \ldots \Big]}{ \prod_{\alpha_i}\,\prod_{j,i}\,\ch{(\s^j_\alpha - \s^i_{\alpha_i})}\, \ch{(\s' - \sum_{\alpha_i}\, \tr \vec \s_{\alpha_i})} \, \ch{(\s' - \tr \vec \s_\alpha + \sum_a \frac{Q_a}{N_a}\, \tr \vec\s_a)} } \nn \\
&= \int \, d\s' \, [d \vec \s_\alpha] \, \frac{e^{2\pi i \eta_\alpha\, (\s' + \sum_{\alpha_i}\, \tr \vec \s_{\alpha_i}  -\tr \vec \s_\alpha)}\, Z_{\rm vec}(\vec \s_\alpha)\, \Big[ \ldots \Big]}{\prod_{\alpha_i}\,\prod_{j,i}\,\ch{(\s^j_\alpha - \s^i_{\alpha_i})}\, \ch{(\s')} \, \ch{(\s' - \tr \vec \s_\alpha +\sum_{\alpha_i}\, \tr \vec \s_{\alpha_i}+ \sum_a \frac{Q_a}{N_a}\, \tr \vec\s_a)}} \nn \\
&= Z^{(\CT^\vee)}(\eta^\vee_{\alpha-1}=\eta_{\alpha-1}+ \eta_\alpha, \eta^\vee_\alpha=- \eta_\alpha, \eta^\vee_{\alpha+1}= \eta_{\alpha+1}+ \eta_\alpha, \eta^\vee = \eta_\alpha, \ldots),
\end{align}
where for the third equality, we have redefined the integration variable $\s'$. The charge vector $\vec Q'$ associated with the quiver 
$\CT^\vee$ can be read off from the third equality and seen to agree with \eref{charge-3}. The extension to the $P > 1$ case is straightforward.

\section{Mutations $I'$ and $II$} \label{app: mutation34}

Mutation $I'$ acts on a $U(N_\alpha)$ gauge node with balance parameter $e_\alpha =1$ 
and a single Abelian hypermultiplet which is charged $N_\alpha$ under the $U(N_\alpha)$ node and has charges $\{ Q_a\}$ under the other unitary gauge groups. 
This mutation is obtained by using the duality $\CD^N_{2N+1,1}$. The mutation deletes the Abelian hyper, replaces the $U(N_\alpha)$ node with a $U(N_\alpha +1)$ node, and ungauges a specific $U(1)$ symmetry of the quiver. Let $J_i$ denote the generator corresponding to the central $U(1)$ subgroup of the gauge group $U(N_i)$. 
The particular $U(1)$ symmetry generator to be ungauged is then given as 
\be
J_G \equiv \sum_a\, \Big(\frac{Q_a}{N_a}\Big)\, J_a  + \sum_i\, J_{\alpha_i}  - J_\alpha, 
\ee
where the first sum extends over all the gauge nodes (aside from the $U(N_\alpha)$ node) under which the Abelian hypermultiplet in 
$\CT$ is charged, and the second sum extends over all the gauge nodes which are connected to $U(N_\alpha)$ by bifundamental 
hypers. In the special case where the Abelian hyper is only charged under the latter gauge nodes with charges $\{-N_i\}$, the ungauging 
operation gives an $SU(N_\alpha +1)$, and the $\CO_{I'}$ operation reduces to the inverse of the operation $\CO_I$. 
The ungauging operation with respect to $\fru(1)_G$ is denoted by $``\Bigg/ U(1)"$ in the quiver.

\begin{figure}[htbp]
\begin{tabular}{ccc}
\scalebox{0.7}{\begin{tikzpicture}
\node[] (1) at (1,0){};
\node[] (100) at (0,0){};
\node[unode] (2) at (2,0){};
\node[unode] (3) at (4,0){};
\node[unode] (4) at (6,0){};
\node[unode] (51) at (2,2){};
\node[] (52) at (0,2){};
\node[] (53) at (1,2){};
\node[unode] (61) at (6,2){};
\node[] (62) at (7,2){};
\node[] (63) at (8,2){};
\node[] (5) at (7,0){};
\node[] (200) at (8,0){};
\node[cross, red] (6) at (4,0.5){};
\draw[-] (1) -- (2);
\draw[-] (2)-- (3);
\draw[-] (3) -- (4);
\draw[-] (4) --(5);
\draw[-] (3) --(51);
\draw[-] (3) --(61);
\draw[-, dotted] (1) -- (100);
\draw[-,dotted] (5) -- (200);
\draw[-, dotted] (52) -- (53);
\draw[-] (51) -- (53);
\draw[-, dotted] (62) -- (63);
\draw[-] (61) -- (62);
\draw[dotted, thick, blue] (0,1.5)--(1,1.5);
\draw[-, thick, blue] (1,1.5)--(2,1.5);
\draw[-, thick, blue] (2)--(2,1.5);
\draw[-, thick, blue] (3)--(4,1.5);
\draw[-, thick, blue] (4)--(6,1.5);
\draw[-, thick, blue] (2,1.5)--(4,1.5);
\draw[-, thick, blue] (4,1.5)--(6,1.5);
\draw[-, thick, blue] (4,1.5)--(51);
\draw[-, thick, blue] (4,1.5)--(61);
\draw[-, thick, blue] (7,1.5)--(6,1.5);
\draw[dotted, thick, blue] (8,1.5)--(7,1.5);
\node[text width=.2cm](11) [below=0.1cm of 2]{$N_{\alpha_2}$};
\node[text width=.2cm](12) at (4.1, -0.5){$N_\alpha$};
\node[text width=.2cm](13) [below=0.1cm of 4]{$N_{\alpha_3}$};
\node[text width=.2cm](15) [above=0.1cm of 51]{$N_{\alpha_1}$};
\node[text width=.2cm](16) [above=0.1cm of 61]{$N_{\alpha_4}$};
\node[text width=.2cm](20) at (4,-1){$(\CT)$};
\end{tikzpicture}} 
&  \scalebox{.7}{\begin{tikzpicture}
\draw[->] (0,0) -- (1.5,0);
\node[text width=0.1cm](29) at (0.7, 0.5) {$\CO_{I'}$};
\node[](30) at (.5, -2) {};
\end{tikzpicture}}
& \scalebox{0.7}{\begin{tikzpicture}
\node[] (1) at (1,0){};
\node[] (100) at (0,0){};
\node[unode] (2) at (2,0){};
\node[unode] (3) at (4,0){};
\node[unode] (4) at (6,0){};
\node[unode] (51) at (2,2){};
\node[] (52) at (0,2){};
\node[] (53) at (1,2){};
\node[unode] (61) at (6,2){};
\node[] (62) at (7,2){};
\node[] (63) at (8,2){};
\node[] (5) at (7,0){};
\node[] (200) at (8,0){};
\draw[-] (1) -- (2);
\draw[-] (2)-- (3);
\draw[-] (3) -- (4);
\draw[-] (4) --(5);
\draw[-] (3) --(51);
\draw[-] (3) --(61);
\draw[-, dotted] (1) -- (100);
\draw[-,dotted] (5) -- (200);
\draw[-, dotted] (52) -- (53);
\draw[-] (51) -- (53);
\draw[-, dotted] (62) -- (63);
\draw[-] (61) -- (62);
\node[text width=.2cm](11) [below=0.1cm of 2]{$N_{\alpha_2}$};
\node[text width=1.5cm](12) at (4.1, -0.5){$N_\alpha +1$};
\node[text width=.2cm](13) [below=0.1cm of 4]{$N_{\alpha_3}$};
\node[text width=.2cm](15) [above=0.1cm of 51]{$N_{\alpha_1}$};
\node[text width=.2cm](16) [above=0.1cm of 61]{$N_{\alpha_4}$};
\node[text width=0.3 cm](10) at (8.5,0){$\Bigg/ U(1)$};
\node[text width=.2cm](20) at (4,-1){$(\CT^\vee)$};
\end{tikzpicture}}
\end{tabular}
\caption{The operation $\CO_{I'}$.}
\label{fig: Mutations I'}
\end{figure}

Given a $U(N_\alpha)$ gauge node with balance parameter $e_\alpha =0$ and $P \geq 1$ Abelian hypermultiplets which are charged 
$N_\alpha$ under the $U(N_\alpha)$ node and have charges $\{ Q^l_a\}_{l=1,\ldots,P}$ under the other unitary gauge groups, we can define a 
mutation $II$ at the node denoted by $\CO_{II}$. Under this mutation, which is obtained by using the duality $\CD^N_{2N,P}$, the gauge and 
flavor nodes remain the same. The $P$ Abelian hypermultiplets, with charge vectors $\vec Q^l= (Q^l_1, \ldots, Q^l_{\alpha_1}, Q^l_{\alpha_2}, N_\alpha, Q^l_{\alpha_3}, Q^l_{\alpha_4}, \ldots, Q^l_L)$ for $l=1,\ldots,P$ and $L$ denoting the total number of nodes in the quiver, are mapped to $P$ Abelian hypermultiplets with charge vectors:
 \be
 \vec Q'^l=(-Q^l_1, \ldots, -Q^l_{\alpha_1} - N_{\alpha_1}, -Q^l_{\alpha_2} - N_{\alpha_2}, N_\alpha, -Q^l_{\alpha_3} - N_{\alpha_3} , -Q^l_{\alpha_4} - N_{\alpha_4}, \ldots, -Q^l_L).
 \ee
 One can check that this operation squares to an identity operation.

\begin{figure}[htbp]
\begin{tabular}{ccc}
\scalebox{0.7}{\begin{tikzpicture}
\node[] (1) at (1,0){};
\node[] (100) at (0,0){};
\node[unode] (2) at (2,0){};
\node[unode] (3) at (4,0){};
\node[unode] (4) at (6,0){};
\node[unode] (51) at (2,2){};
\node[] (52) at (0,2){};
\node[] (53) at (1,2){};
\node[unode] (61) at (6,2){};
\node[] (62) at (7,2){};
\node[] (63) at (8,2){};
\node[] (5) at (7,0){};
\node[] (200) at (8,0){};
\node[cross, red] (6) at (4,0.5){};
\draw[-] (1) -- (2);
\draw[-] (2)-- (3);
\draw[-] (3) -- (4);
\draw[-] (4) --(5);
\draw[-] (3) --(51);
\draw[-] (3) --(61);
\draw[-, dotted] (1) -- (100);
\draw[-,dotted] (5) -- (200);
\draw[-, dotted] (52) -- (53);
\draw[-] (51) -- (53);
\draw[-, dotted] (62) -- (63);
\draw[-] (61) -- (62);
\draw[dotted, thick, blue] (0,1.5)--(1,1.5);
\draw[-, thick, blue] (1,1.5)--(2,1.5);
\draw[-, thick, blue] (2)--(2,1.5);
\draw[-, thick, blue] (3)--(4,1.5);
\draw[-, thick, blue] (4)--(6,1.5);
\draw[-, thick, blue] (2,1.5)--(4,1.5);
\draw[-, thick, blue] (4,1.5)--(6,1.5);
\draw[-, thick, blue] (4,1.5)--(51);
\draw[-, thick, blue] (4,1.5)--(61);
\draw[-, thick, blue] (7,1.5)--(6,1.5);
\draw[dotted, thick, blue] (8,1.5)--(7,1.5);
\node[text width=.2cm](11) [below=0.1cm of 2]{$N_{\alpha_2}$};
\node[text width=.2cm](12) at (4.1, -0.5){$N_\alpha$};
\node[text width=.2cm](13) [below=0.1cm of 4]{$N_{\alpha_3}$};
\node[text width=.2cm](15) [above=0.1cm of 51]{$N_{\alpha_1}$};
\node[text width=.2cm](16) [above=0.1cm of 61]{$N_{\alpha_4}$};
\node[text width=0.1 cm](10) at (4, 2){\footnotesize{$\vec Q$}};
\node[text width=.2cm](20) at (4, -1){$(\CT)$};
\end{tikzpicture}}
& \scalebox{.7}{\begin{tikzpicture}
\draw[->] (0,0) -- (1.5,0);
\node[text width=0.1cm](29) at (0.7, 0.5) {$\CO_{II}$};
\node[](30) at (.5, -2) {};
\end{tikzpicture}}
& \scalebox{0.7}{\begin{tikzpicture}
\node[] (1) at (1,0){};
\node[] (100) at (0,0){};
\node[unode] (2) at (2,0){};
\node[unode] (3) at (4,0){};
\node[unode] (4) at (6,0){};
\node[unode] (51) at (2,2){};
\node[] (52) at (0,2){};
\node[] (53) at (1,2){};
\node[unode] (61) at (6,2){};
\node[] (62) at (7,2){};
\node[] (63) at (8,2){};
\node[] (5) at (7,0){};
\node[] (200) at (8,0){};
\draw[-] (1) -- (2);
\draw[-] (2)-- (3);
\draw[-] (3) -- (4);
\draw[-] (4) --(5);
\draw[-] (3) --(51);
\draw[-] (3) --(61);
\draw[-, dotted] (1) -- (100);
\draw[-,dotted] (5) -- (200);
\draw[-, dotted] (52) -- (53);
\draw[-] (51) -- (53);
\draw[-, dotted] (62) -- (63);
\draw[-] (61) -- (62);
\draw[dotted, thick, blue] (0,1.5)--(1,1.5);
\draw[-, thick, blue] (1,1.5)--(2,1.5);
\draw[-, thick, blue] (2)--(2,1.5);
\draw[-, thick, blue] (3)--(4,1.5);
\draw[-, thick, blue] (4)--(6,1.5);
\draw[-, thick, blue] (2,1.5)--(4,1.5);
\draw[-, thick, blue] (4,1.5)--(6,1.5);
\draw[-, thick, blue] (4,1.5)--(51);
\draw[-, thick, blue] (4,1.5)--(61);
\draw[-, thick, blue] (7,1.5)--(6,1.5);
\draw[dotted, thick, blue] (8,1.5)--(7,1.5);
\node[text width=.2cm](11) [below=0.1cm of 2]{$N_{\alpha_2}$};
\node[text width=.2cm](12) at (4.1, -0.5){$N_\alpha$};
\node[text width=.2cm](13) [below=0.1cm of 4]{$N_{\alpha_3}$};
\node[text width=.2cm](15) [above=0.1cm of 51]{$N_{\alpha_1}$};
\node[text width=.2cm](16) [above=0.1cm of 61]{$N_{\alpha_4}$};
\node[text width=0.1 cm](10) at (4, 2){\footnotesize{$\vec Q'$}};
\node[text width=.2cm](20) at (4,-1){$(\CT^\vee)$};
\end{tikzpicture}}
\end{tabular}
\caption{The operation $\CO_{II}$ for $P=1$.}
\label{fig: Mutations II}
\end{figure}

The gauge node $U(N_\alpha -1)$ of the theory $\CT^\vee$ in \eref{fig: Mutations III} has a balance parameter $e_\alpha =1$.  One can check that if one implements $\CO_{I'}$ at the gauge node $U(N_\alpha -1)$ of the quiver $\CT^\vee$ (after an appropriate field redefinition in the theory), one gets back the quiver $\CT$. The composition of $\CO_{I'}$ with $\CO_{III}$ therefore gives the identity operation. The inverse of $\CO_I$ is also a special case of an $\CO_{I'}$ operation.

\appendix

\end{document}